\documentclass[conference,a4paper]{IEEEtran}

\ifCLASSINFOpdf
\else
\fi
\usepackage[left=1.57cm,right=1.57cm,top=0.95cm,bottom=2.54cm]{geometry}
\usepackage{listings}
\usepackage{setspace}
\usepackage{graphicx}
\usepackage{tabularx,booktabs}
\usepackage{dingbat}
\usepackage{diagbox}
\usepackage{multirow} 
\usepackage[hyphens]{url}
\usepackage[hidelinks,breaklinks]{hyperref}
\usepackage{xcolor}
\usepackage{amsfonts, amsmath, amsthm, amssymb}
\usepackage{subcaption}
\hypersetup{breaklinks=true}
\usepackage{tikz}
\usepackage{textcomp}
\usepackage{lipsum}
\usepackage{svg}
\IEEEoverridecommandlockouts
\newcommand\copyrighttext{%
  \footnotesize \textcopyright 2026 IEEE.  Personal use of this material is permitted.  Permission from IEEE must be obtained for all other uses, in any current or future media, including reprinting/republishing this material for advertising or promotional purposes, creating new collective works, for resale or redistribution to servers or lists, or reuse of any copyrighted component of this work in other works.}
\newcommand\copyrightnotice{%
\begin{tikzpicture}[remember picture,overlay]
\node[anchor=south,yshift=10pt] at (current page.south) {\fbox{\parbox{\dimexpr\textwidth-\fboxsep-\fboxrule\relax}{\copyrighttext}}};
\end{tikzpicture}%
}
\usepackage{graphicx}
\graphicspath{{figures/}}
\begin{document}

%
\title{Field Test of 5G New Radio (NR) UL-MIMO and UL-256QAM for HD Live-Streaming}

\author{
    \IEEEauthorblockN{Kasidis Arunruangsirilert}
    \IEEEauthorblockA{Department of Computer Science and Communications Engineering, Waseda University, Tokyo, Japan
    \\kasidis@katto.comm.waseda.ac.jp}
}
%

\maketitle

\copyrightnotice
\setstretch{0.95}
\begin{abstract}


The exponential growth of User-Generated Content (UGC), especially High-Definition (HD) live video streaming, places a significant demand on the uplink capabilities of mobile networks. To address this, the 5G New Radio (NR) standard introduced key uplink enhancements, including Uplink Multi-Input Multi-Output (UL-MIMO) and Uplink 256QAM, to improve throughput and spectral efficiency. However, while the benefits of these features for raw data rates are well-documented, their practical impact on real-time applications like live-streaming is not yet well understood. This paper investigates the performance of UL-MIMO and UL-256QAM for HD live-streaming over a commercial 5G network using the Real-Time Messaging Protocol (RTMP). To ensure a fair assessment, we conduct a comparative analysis by modifying the modem firmware of commercial User Equipment (UE), allowing these features to be selectively enabled and disabled on the same device. Performance is evaluated based on key metrics, including dropped video frames and connection stability. Furthermore, this study analyzes 5G Radio Frequency (RF) parameters to quantify the spectral efficiency impact, specifically examining metrics derived from the Channel State Information (CSI) framework, including Reference Signal Received Power (CSI-RSRP), Reference Signal Received Quality (CSI-RSRQ), and Signal-to-Interference-plus-Noise Ratio (CSI-SINR).

\end{abstract}

\begin{IEEEkeywords}
5G New Radio (NR), Uplink 256QAM, Uplink MIMO, Live-Streaming, Radio Access Network
\end{IEEEkeywords}


\setstretch{0.947}

%
\IEEEpeerreviewmaketitle

\section{Introduction}


The emergence of 5G New Radio (NR) has enabled new applications such as massive IoT, Smart City, cloud gaming, Autonomous Vehicles, and Virtual/Augmented Reality (VR/AR) \cite{10373900, 10574348}. A key use case is transmitting of User Generated Content (UGC), a trend that has grown since the early 2010s \cite{4801529}. The increasing popularity of social media platforms, which encourage users to create and share multimedia content like HD live video \cite{ericsson_2016, 10951237}, has presented new challenges for mobile networks traditionally optimized for downlink traffic. During the 4G Long-Term Evolution (LTE) era, mobile network traffic was predominantly downlink-heavy, with most users consuming existing internet content rather than producing them. The recent rise of multimodal Generative AI (GenAI), capable of processing images, audio, and video, has further increased uplink transmissions as users attach multimedia payloads to their prompts \cite{ericsson_2024}. This evolving usage pattern is causing mobile traffic to become more balanced as we progress toward the 6G era.


To address the growing demand for high-performance uplink, 3GPP has introduced several new features in the 5G Standard \cite{3GPP_38-101-1}, including Uplink MIMO (UL-MIMO) and Uplink 256QAM (UL-256QAM). Unlike conventional 4G LTE User Equipment (UE), which typically has only one uplink antenna, 5G UEs may possess more than one. This allows for the adoption of UL-MIMO, where a UE with two antennas can use spatial multiplexing to double its maximum uplink throughput in ideal signal conditions or operate in transmit diversity mode to enhance connection reliability in weaker signal conditions. A previous study found that in a real-world commercial network, UL-MIMO-enabled UEs deliver an average of 19.8\% higher throughput on moving trains in urban areas, with up to a 33.5\% improvement observed in more favorable signal conditions on the C-Band spectrum \cite{10118777}. However, since dual transmission antenna implementation is not mandatory under the 3GPP standard, many UE manufacturers omit this optional feature for design simplicity and cost-saving.


Alternatively, Uplink 256QAM is simpler to implement as it requires no additional antennas on the UE side. It works by assigning Modulation Coding Scheme (MCS) Index Table 2 (\textit{qam256}) for uplink transmission instead of MCS Index Table 1 (\textit{qam64}), as standardized in \cite{3GPP_38-214}. This can improve spectral efficiency by a maximum of 33.3\% by allowing each symbol to carry 8 bits instead of the 6 bits permitted by 64QAM. The effectiveness of Uplink 256QAM, however, is highly dependent on the signal condition, requiring a Signal-to-Noise Ratio of over 30 dB to maintain stable communication, which is difficult to achieve in a real-world commercial network environment. In our previous work \cite{10570635}, we confirmed this, finding that enabling UL-256QAM resulted in only 20\% of resource blocks being modulated using 256QAM on a passive antenna network, yielding an uplink throughput gain of just 8.22\%. \looseness=-1


While both studies highlight the advantages of UL-MIMO and UL-256QAM in enhancing uplink throughput and spectral efficiency, their impacts on real-time applications such as live-streaming remain largely unknown. Since these 5G features are expected to be key enablers for Ultra High-Definition (UHD) live-streaming, understanding their impact is crucial for planning Next-Generation Radio Access Networks (NG-RANs) to support emerging use cases. In this paper, the performance impact of UL-MIMO and UL-256QAM on the live-streaming of HD video is investigated. By modifying the modem firmware, both features can be enabled and disabled on the same User Equipment, ensuring a fair comparison. Videos were transmitted to an RTMP server via a commercial 5G UE and network, where performance was evaluated in terms of dropped video frames and the number of disconnections. Furthermore, 5G RF parameters were analyzed to understand the spectral efficiency impact of each feature, including key metrics from the Channel State Information (CSI) framework like CSI-RSRP, CSI-RSRQ, and CSI-SINR, which is crucial for RAN capacity planning.

\section{Field Test Setup}

\subsection{Uplink MIMO Evaluation Environment}

For the evaluation of Uplink MIMO, a User Equipment (UE) with two transmission antennas is required. Therefore, a KDDI au Speed Wi-Fi HOME 5G L13, based on the ZTE MC888 Ultra, equipped with the Qualcomm Snapdragon X65 5G RF Modem, was used \cite{qualcomm_X65}. The modem's diagnostic port was exposed via an internal service mode, which permitted firmware modifications to limit the capability to a single transmission (Single-Tx) as needed. To collect RF parameters, the device was connected to a laptop running \textit{AirScreen}, a professional network drive test tool. Before each trial, \textit{UECapabilityInformation} packets were verified to ensure the UE's capability was configured as expected. The field test for UL-MIMO was conducted on the network of a Japanese Mobile Network Operator (MNO), SoftBank, using a commercial plan and SIM card.

Since the UE used for this experiment supports UL-MIMO only on Time Division Duplexing (TDD) frequency bands, the JR Yamanote line in central Tokyo was chosen for the test, as it has good coverage of SoftBank's 5G Band n77 (3.4 GHz). To address potential result variability from the UE randomly connecting to SoftBank's multiple n77 carriers—each with different path loss, coverage, and bandwidth, such as 3.4 GHz (40 MHz Bandwidth - NR-ARFCN 627360), 3.5 GHz (40 MHz Bandwidth - NR-ARFCN 638112), and 3.9 GHz (100 MHz Bandwidth - NR-ARFCN 660768)—the modem firmware was modified. This change locked the UE to 5G Standalone Mode on the target NR-ARFCN of 627360, which is SoftBank's lowest frequency n77 carrier with the most base stations. Each trial involved looping the Yamanote Line once on a moving train with an average speed of 28 km/h according to GPS data, taking about 70 minutes to complete. The RF conditions are summarized in Table \ref{tab-RFCondition} for reference purposes. It should be noted that UL-256QAM was also enabled for these test scenarios. The collected RF data were then analyzed using AirScreen, the same software that was utilized to collect the logs. \looseness=-1

\subsection{Uplink 256QAM Evaluation Environment}

To evaluate Uplink 256QAM, the Japanese Samsung Galaxy S22 (SC-52C), which is equipped with the same Qualcomm Snapdragon X65 5G RF Modem, served as the User Equipment (UE). Since this smartphone runs the Android operating system, \textit{Network Signal Guru (NSG)}, the Android version of \textit{AirScreen}, was installed on the phone to collect RF parameters. The same commercial plan and SIM card were used for this test. By accessing Samsung's service mode menu, the UL-256QAM capability could be turned off as needed for comparison. This study focused on two of SoftBank's Frequency Division Duplexing (FDD) spectrums, n3 (1800 MHz - 15 MHz Bandwidth) and n28 (700 MHz - 10 MHz Bandwidth), because MNOs globally are refarming FDD bands from LTE service to 5G NR, which enables UL-256QAM on such spectrum. It is important to note that the n28 band is configured purely as 5G NR, whereas the n3 band employs Dynamic Spectrum Sharing (DSS), permitting the simultaneous use of the spectrum for both LTE and 5G NR services. To ensure each band could be tested independently, the UE was locked to a specific frequency using the forcing function within the NSG application.

The use of these FDD bands facilitates an investigation into not only the impact of UL-256QAM but also the advantages of migrating from LTE to 5G NR, since both Radio Access Technologies use Single-carrier FDMA (SC-FDMA) for the uplink. The experiment was conducted on Tokyu Den-en-toshi Line, excluding the underground section (Futako-tamagawa to Minami-machida Grandberry Park) due to densely deployed SoftBank FDD 5G NR base stations along the line. Each trial took approximately 30 minutes to complete on a moving train with an average speed of 51.5 km/h based on GPS data. Similar to the UL-MIMO evaluation, RF conditions can be found in Table \ref{tab-RFCondition}. Since the smartphone has only a Single Tx antenna, UL-MIMO is not supported and hence not enabled in these test scenarios. Following the trials, the RF logs from \textit{Network Signal Guru (NSG)} were analyzed using \textit{AirScreen} on a PC. 

\subsection{Live-Streaming Configuration}


To understand the behavior of live-streaming applications utilizing the RTMP protocol, an RTMP server was hosted on a server in Tokyo, Japan, using SRS Media Server (v6.0.155). For the client, Open Broadcaster Software (OBS) Studio (v30.1.1) was installed on a laptop equipped with an Intel Wi-Fi 6 AX200 adapter. The laptop was connected to the network via Wi-Fi, either directly to a KDDI au Speed Wi-Fi HOME 5G L13 router or through the built-in tethering function of a Samsung Galaxy S22 Ultra.

In OBS, the live video was encoded using the H.264/AVC codec via the AMD Advanced Media Framework (AMF) hardware encoder, with the rate control mode set to Constant Bit Rate (CBR) to provide a stable uplink load for isolating network performance. Furthermore, the ``\textit{Dynamically change bitrate to manage congestion (Beta)}" feature was not enabled, as it is unsupported by the AMD AMF hardware encoder used in this study. Following live encoding recommendations from YouTube and Twitch \cite{google, twitch}, the keyframe interval was set to two seconds, and the maximum number of consecutive B-frames was set to two. Based on preliminary experiments, target bitrates of 5000 kbps and 3000 kbps were selected for the UL-MIMO and UL-256QAM evaluations, respectively. The video was transmitted to the server using the RTMP protocol, and performance was evaluated by examining the OBS logs for the number of disconnections and dropped frames. \looseness=-1

\begin{table}[!tbp]
\setstretch{0.75}
\caption{Summary of RF Conditions by Test Case}
\vspace{-1.7mm}
\centering
\label{tab-RFCondition}
\resizebox{8.5cm}{!}{\begin{tabular}{@{}cccccccccc@{}}
\toprule
Freq&DL&UL&SSB&BW&MCS&Tx&CSI&CSI&CSI \\
Band&Freq&Freq&ARFCN&(MHz)&Table&Conf.&RSRP&RSRQ&SINR\\
&(MHz)&(MHz)&&&&&(dBm)&(dB)&(dB)\\\midrule
\multicolumn{10}{c}{Uplink MIMO Evaluation}\\\midrule
\multirow{2}{*}{n77}&\multirow{2}{*}{3420}&\multirow{2}{*}{3420}&\multirow{2}{*}{627360}&\multirow{2}{*}{40}&\multirow{2}{*}{\textit{qam256}}&1-Tx&-92.28&-15.43&8.76\\
&&&&&&2-Tx&-91.71&-15.53&8.58\\\midrule
\multicolumn{10}{c}{Uplink 256QAM Evaluation}\\\midrule
\multirow{2}{*}{n28}&\multirow{2}{*}{798}&\multirow{2}{*}{743}&\multirow{2}{*}{159630}&\multirow{2}{*}{10}&\textit{qam64}&\multirow{2}{*}{1-Tx}&-86.79&-15.33&6.34\\
&&&&&\textit{qam256}&&-89.60&-15.33&6.22\\
\midrule
\multirow{2}{*}{n3}&\multirow{2}{*}{1852.5}&\multirow{2}{*}{1757.5}&\multirow{2}{*}{369870}&\multirow{2}{*}{15}&\textit{qam64}&\multirow{2}{*}{1-Tx}&-90.19&-15.46&8.96\\
&&&&&\textit{qam256}&&-90.79&-15.60&8.78\\
\bottomrule
\end{tabular}}
\vspace{-5mm}
\end{table}

\section{Result and Analysis}

\subsection{Uplink MIMO}

\begin{figure}[t!]
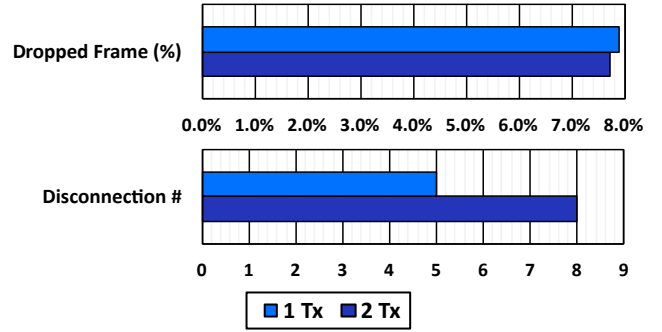

\vspace{2mm}
\centering\includesvg[width=0.95\linewidth,inkscapelatex=false]{OBS-ULMIMO.svg}\\[2mm]
\centering\includesvg[width=0.25\linewidth,inkscapelatex=false]{ULMIMO_Legend.svg}

\caption{\textbf{UL-MIMO}: Live Encoding Performance}
\label{fig:OBS-ULMIMO}
\vspace{-7mm}
\end{figure}

\begin{figure}[t!]
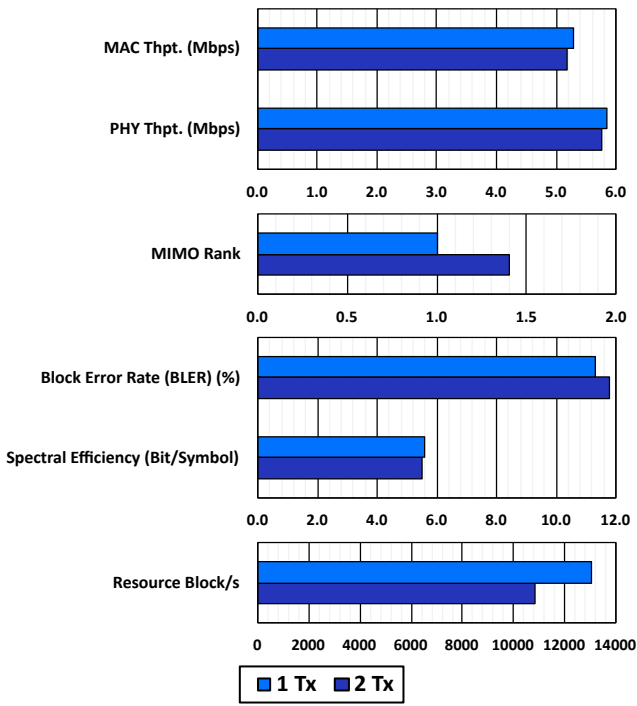

\centering\includesvg[width=0.95\linewidth,inkscapelatex=false]{RF-ULMIMO.svg}\\[2mm]
\centering\includesvg[width=0.25\linewidth,inkscapelatex=false]{ULMIMO_Legend.svg}

\caption{\textbf{UL-MIMO}: RF Parameters}
\vspace{-5mm}
\label{fig:RF-ULMIMO}
\end{figure}

\begin{figure}[t!]
\centering\includesvg[width=0.90\linewidth,inkscapelatex=false]{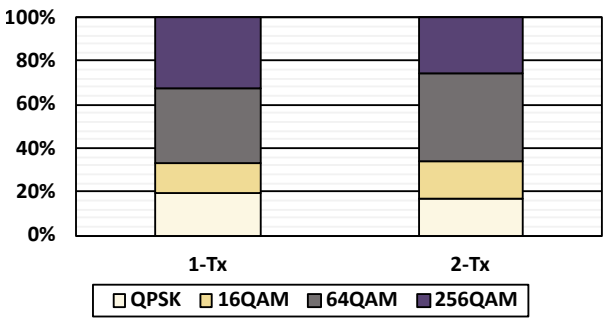}

\caption{\textbf{UL-MIMO}: Uplink Modulation Utilization (\%)}
\vspace{-6.5mm}
\label{fig:QAM-ULMIMO}

\end{figure}


Figure \ref{fig:OBS-ULMIMO} presents the dropped frame percentage and the number of disconnections from the RTMP server. The results indicate that UL-MIMO reduces the dropped frame percentage by a marginal 0.1\%. However, this slight improvement is accompanied by a significant increase in connection instability. When UL-MIMO was enabled, the OBS client re-established its connection to the RTMP server eight times, a substantial increase compared to five times for the single-transmit configuration. This instability arises because a standard UE, operating under Power Class 3, is limited to a maximum transmission power of 23 dBm. With UL-MIMO enabled, this power budget is divided between two transmission antennas, effectively limiting each MIMO chain to a maximum of 20 dBm. Although a UL-MIMO-capable UE can utilize transmission diversity to compensate for this reduced power per chain, a single, more powerful transmission is more effective at overcoming propagation loss in poor channel conditions, thus providing a more stable connection. To address this limitation, 3GPP has standardized High Power User Equipment (HPUE), which allows certified UEs in Power Class 2 (PC2) and Power Class 1.5 (PC1.5) to transmit at higher powers of 26 dBm and 29 dBm, respectively. This restores the power per antenna to levels comparable to the single-transmit case, thereby mitigating the negative effects of power division across multiple antenna chains \cite{3GPP_38-101-1}. The performance impact of HPUE on real-time applications is beyond the scope of this study and will be investigated in future work. \looseness=-1

Figure \ref{fig:RF-ULMIMO} illustrates the RF parameters obtained from \textit{AirScreen}. Due to the use of a Constant Bit Rate (CBR) rate control mode, there was no significant difference in the MAC layer or Physical (PHY) layer throughput between the two trials. However, enabling UL-MIMO resulted in a slight increase in the Block Error Rate (BLER) by 0.49\%, indicating a higher re-transmission probability, which could potentially increase latency. Despite this, the primary advantage of UL-MIMO is its ability to utilize spatial multiplexing, allowing for the transmission of two independent data streams. The effectiveness of this operation is quantified by the MIMO Rank, which indicates the number of data streams being transmitted. The results show that the UL-MIMO configuration achieved an average MIMO Rank of 1.41, a significant improvement over the fixed rank of 1.00 in the single-transmit configuration. This enhancement led to a 17.05\% reduction in the number of uplink resource blocks consumed for the Physical Uplink Shared Channel (PUSCH), signifying a substantial improvement in spectral efficiency.

Further analysis of spectral efficiency is presented in Figure \ref{fig:QAM-ULMIMO}, which details the utilization percentage of each modulation scheme for PUSCH. While enabling UL-MIMO led to a 7.3\% decrease in the use of 256QAM, the availability of transmission diversity compensated for this in medium-quality channel conditions. Approximately 3\% of resource blocks were modulated using 16QAM (4 bits/symbol) instead of QPSK (2 bits/symbol), which partially offset the spectral efficiency loss from the reduced 256QAM utilization. In summary, although enabling UL-MIMO may lead to increased disconnections in challenging RF environments, it improves spectral efficiency by 17.05\%. This allows for more effective use of the frequency spectrum, which can enhance cell capacity in congested scenarios like mass gathering and live concerts. The adoption of HPUE on UL-MIMO-capable UEs is expected to mitigate connection stability issues and improve the overall Quality-of-Experience (QoE) for live video transmission applications.


\subsection{Uplink 256QAM}

\begin{figure}[t!]
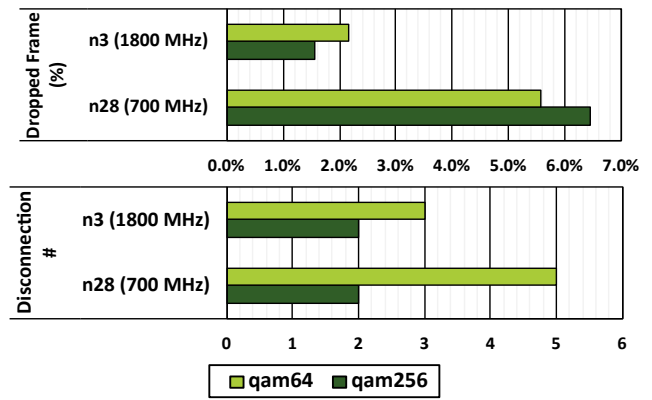

\centering\includesvg[width=0.95\linewidth,inkscapelatex=false]{OBS-256QAM.svg}\\[2mm]
\centering\includesvg[width=0.35\linewidth,inkscapelatex=false]{256QAM_Legend.svg}

\caption{\textbf{UL-256QAM}: Live Encoding Performance}
\vspace{-7mm}
\label{fig:OBS-256QAM}
\end{figure}

\begin{figure}[t!]
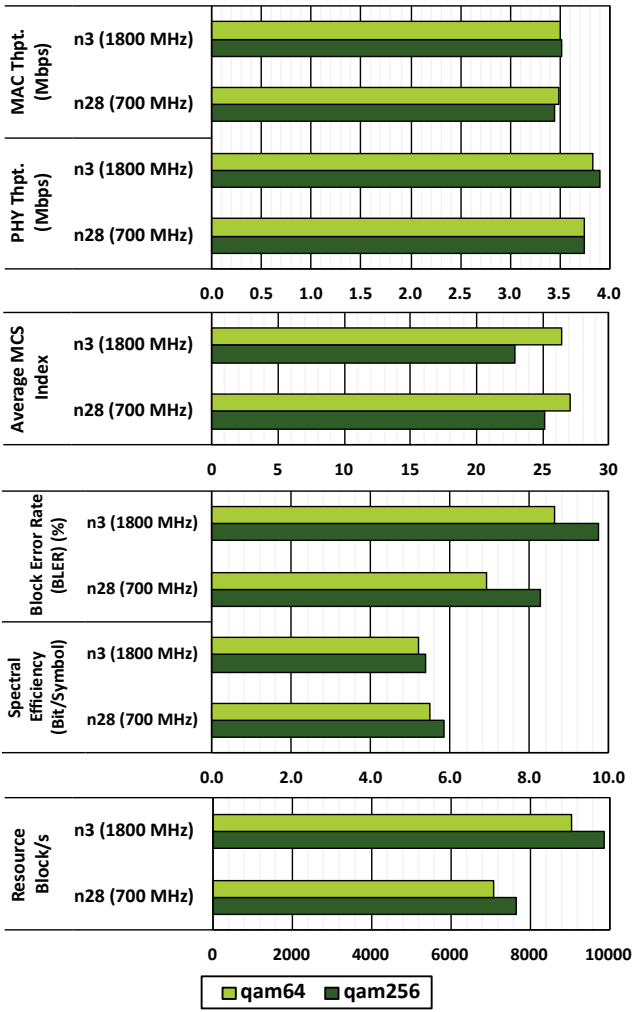

\centering\includesvg[width=0.95\linewidth,inkscapelatex=false]{RF-256QAM.svg}\\[2mm]
\centering\includesvg[width=0.35\linewidth,inkscapelatex=false]{256QAM_Legend.svg}

\caption{\textbf{UL-256QAM}: RF Parameters}
\vspace{-5mm}
\label{fig:RF-256QAM}
\end{figure}

\begin{figure}[t!]
\centering\includesvg[width=0.85\linewidth,inkscapelatex=false]{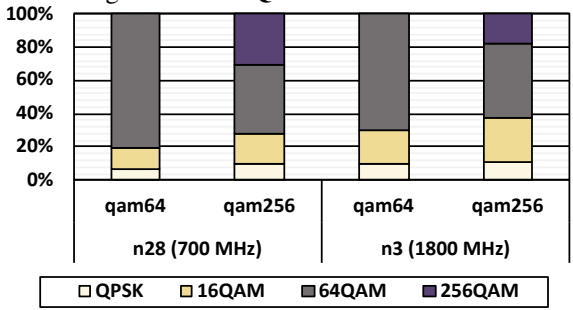}

\caption{\textbf{UL-256QAM}: Uplink Modulation Utilization (\%)}
\label{fig:QAM-256QAM}
\vspace{-7.5mm}
\end{figure}


Figure \ref{fig:OBS-256QAM} presents the dropped frame percentage and the number of disconnections from the RTMP server on frequency bands n3 (1800 MHz) and n28 (700 MHz), comparing performance with UL-256QAM enabled (\textit{qam256} MCS Table) versus disabled (\textit{qam64} MCS Table). The results for dropped frames were mixed; enabling UL-256QAM decreased the dropped frame rate by 0.6\% on the higher frequency band n3 but increased it by 0.8\% on the lower frequency band n28. However, a notable improvement in connection stability was observed. Enabling UL-256QAM reduced the number of connection re-establishments from three on band n3 and five on band n28 to a consistent two across both bands, suggesting a more stable user experience.


Figure \ref{fig:RF-256QAM} shows the RF parameters logged by \textit{Network Signal Guru (NSG)}. As in the Uplink MIMO experiments, the use of a Constant Bit Rate (CBR) mode resulted in no significant throughput difference between scenarios. However, enabling UL-256QAM had a detrimental effect on link reliability at the physical layer, evidenced by a significant increase in the Block Error Rate (BLER). The BLER increased from 6.92\% to 8.27\% for band n3 and from 8.62\% to 9.72\% for band n28. This degradation led to a decrease in spectral efficiency, as the number of uplink resource blocks consumed by the UE increased by 9.19\% and 8.43\% for bands n3 and n28, respectively.\looseness=-1


As detailed in Figure \ref{fig:QAM-256QAM}, which illustrates the utilization of each PUSCH modulation scheme. When the \textit{qam256} MCS table was enabled, 17.7\% and 30.8\% of resource blocks were modulated using 256QAM on bands n3 and n28, respectively. However, the utilization of lower-order modulations like QPSK and 16QAM also increased by approximately 5\% each, which negated the potential spectral efficiency gains. This outcome is a consequence of the MCS table implementation in the 5G standard, which permits a maximum of 32 entries. The \textit{qam256} table accommodates the higher modulation by removing intermediate target code rates for lower-order modulations. For instance, the \textit{qam64} table defines ten, seven, and twelve target code rates for QPSK, 16QAM, and 64QAM, respectively, whereas the \textit{qam256} table reduces these to five, six, and nine. This reduced granularity impairs link-adaptation in suboptimal conditions. Although some RAN vendors offer proprietary solutions to dynamically switch between MCS tables based on signal quality, such features are beyond the scope of this work.


In conclusion, while enabling UL-256QAM can improve channel capacity and result in a more stable application-layer connection, it negatively impacts physical layer performance. The UE consumes significantly more resource blocks and exhibits a higher BLER, offsetting the spectral efficiency gains from 256QAM. Therefore, MNOs should carefully evaluate channel quality and cell capacity before deploying this feature. Furthermore, future Next-Generation RAN (NG-RAN) standards should consider expanding the number of MCS table entries to better support higher modulation schemes such as 1024QAM.

\section{Conclusions and Future Work}


In this paper, the performance of Uplink MIMO (UL-MIMO) and UL-256QAM for live HD video streaming over a commercial 5G Standalone (SA) network were investigated. The results show a significant trade-off between improving spectral efficiency and maintaining connection stability for real-time applications. Enabling UL-MIMO demonstrated a substantial improvement in spectral efficiency by reducing resource block consumption, which is highly beneficial for increasing cell capacity in congested environments. However, this gain came at the cost of reduced connection stability, evidenced by a significant increase in RTMP disconnections when UL-MIMO was active. This instability is attributed to the division of the User Equipment's (UE's) 23 dBm power budget across two antennas, which can impair performance in poor channel conditions.


Conversely, enabling UL-256QAM improved application-layer stability by reducing the number of disconnections. However, this benefit was offset by degraded physical layer performance, as the results show a significantly increased Block Error Rate (BLER) and an 8-9\% increase in resource block consumption. The degradation in physical layer link stability is due to the limited number of entries in the standardized \textit{qam256} Modulation and Coding Scheme (MCS) table, which impairs link-adaptation capabilities and negates potential spectral efficiency gains. These results indicate that while these 5G features can enhance network performance, their application to live streaming requires careful consideration of the deployment environment, number of users, cell planning, and overall channel quality.


For future work, an investigation into High-Power User Equipment (HPUE) is needed to determine if its higher transmission power can mitigate the connection instability observed with UL-MIMO, thereby allowing applications to benefit from its spectral efficiency gains without compromising reliability . Additionally, future Next-Generation RAN (NG-RAN) standards should address the limitations of the current MCS Table design, which impairs link-adaptation capabilities, to better accommodate higher-order modulations such as 1024QAM as networks evolve toward 6G . Finally, to generalize the findings beyond the specific context of this study, urban commuter rail lines in Tokyo, subsequent research should encompass a wider variety of environments, including diverse geographies, mobility scenarios, network operators, and real-time protocols such as SRT or WebRTC.

\section*{Acknowledgement}

This paper is supported by the Japan Science and Technology Agency (JST) CRONOS Grant Number JPMJCS25N2. Additionally, the author would like to express their gratitude to \textbf{PEI Xiaohong} of \textit{Qtrun Technologies} for providing \textit{Network Signal Guru (NSG)} and \textit{AirScreen}, the cellular network drive test software used for result collection and analysis in this research.




%
\setstretch{1}
\Urlmuskip=0mu plus 1mu\relax
\bibliographystyle{IEEEtran}
\bibliography{b_reference}

\end{document}